\documentclass[10pt,emptycopyrightspace]{ewsn-proc}

\usepackage{graphicx}
\usepackage{balance}
\usepackage{comment}
\usepackage{bm}
\usepackage{color}
\usepackage{subfigure}
\usepackage{amsmath}
\usepackage{tabularx}
\usepackage{graphicx}
\usepackage{booktabs}
\usepackage{amsfonts}

\numberofauthors{1}
\author{
    Haifeng Jia$^*$, Yichen Wei$^*$, Zhan Wang, Jiani Jin, Haorui Li, and Yibo Pi$^\dagger$\\
    {Shanghai Jiao Tong University, China}\\
    {\{ethanjia, ethepherein, ababba, jinjiani, haorui.li, yibo.pi\}@sjtu.edu.cn}\\%
}

\title{Efficient Interference Graph Estimation via Concurrent  Flooding}

\begin{document}

\maketitle

\def\thefootnote{*}\footnotetext{Equal contribution}
\def\thefootnote{$^\dagger$}\footnotetext{Corresponding author}

\begin{abstract}
Traditional wisdom for network management allocates network resources separately for the measurement and data transmission tasks. Heavy measurement tasks may take up resources for data transmission and significantly reduce network performance. It is therefore challenging for interference graphs, deemed as incurring heavy measurement overhead, to be used in practice in wireless networks. To address this challenge in wireless sensor networks, we propose to use power as a new dimension for interference graph estimation (IGE) and integrate IGE with concurrent flooding such that IGE can be done simultaneously with flooding using the same frequency-time resources. 
With controlled and real-world experiments, we show that it is feasible to efficiently achieve IGE via concurrent flooding on the commercial off-the-shelf (COTS) devices by controlling the transmit powers of nodes. We believe that efficient IGE would be a key enabler for the practical use of the existing scheduling algorithms assuming known interference graphs.

\end{abstract}

%
%

%

\section{Introduction}
  \label{sec:intro}

Interference graphs, depicting the channel conditions between network nodes, are a key element for resource management in wireless networks. Given an interference graph, we can estimate the interference of each node to other nodes in the network and allocate network resources (e.g., power, time, and frequency) in a proper manner. Extensive resource allocation schemes have been proposed for wireless sensor networks assuming the availability of the interference graphs~\cite{wsn_interference_3, wsn_interference_1, wsn_interference_2}. However, compared to its usage, the advances in the efficient measurement of interference graphs are still lacking, which greatly hinders the practical use of existing resource allocation schemes in wireless sensor networks. Efficient interference graph estimation (IGE) has great potential in unleashing the power of resource scheduling in wireless sensor networks.

Traditional wisdom for network management is to allocate  network resources (i.e., time and frequency) separately for the measurement and data transmission tasks. In other words, the measurement and data transmission tasks have to compete for the time and frequency resources. As a result, on the same channel, the measurement tasks, if not done efficiently, will significantly reduce the opportunities for data transmission. Unfortunately, under the traditional wisdom for network management, IGE inherently incurs high measurement overhead: a $N$-node network includes $O(N^2)$ total links, which consumes at least $N$ slots for measurement, where one node sends a reference signal to all the other nodes for channel estimation in each slot~\cite{ige_traditional_survey, ige_traditional_wisdom}. Such high measurement overhead makes the traditional wisdom not scalable for large networks, not to mention network dynamics requiring frequent updates on the interference graph. To reduce the measurement overhead, another popular approach is to characterize interference by modelling based on the propagation environment, where model parameters are pre-determined and no measurement is needed. Similarly, several recent works have
assumed the existence of a direct mapping between node attributes, e.g., geolocation, and resource scheduling decisions, which can be learned from a large amount of network layouts by deep learning~\cite{ige_spatial_dl,ige_embedding}. Without active measurements, these approaches lack the ability to track network dynamics.

In this paper, we propose to integrate measurement tasks into data transmission tasks, contrary to the traditional wisdom of separating them, such that the measurement tasks can be conducted together with the data transmission tasks, without reserving time resources solely for conducting the measurement tasks. To achieve this, our core insight is to decompose the superposition of the received signal strengths from multiple concurrent senders into the signal strengths of individually attenuated transmit powers in the data transmission tasks. Specifically, our approach assumes \emph{the linearity of received power}: the received power of a listener is a linear combination of the channel gains and the transmit powers of the senders. By varying the transmit powers of the senders, we can obtain different received powers of the listener. Then, the channel gains can be obtained by solving a group of equations, if the received powers of the listener and the transmit powers of the senders are all known. 

Although the core insight is intriguing, it is difficult to find a suitable data transmission task for integration, which should at least satisfy the following two requirements: 1) it needs to be conducted in a well-synchronized network, such that the signal strength measured by the listener in a period can be matched to the set of the senders transmitting in the same period; 2) the received powers from concurrent senders are additive for the linearity assumption of received power to be true. The second one is challenging to satisfy, requiring not only strictly-synchronized concurrent senders for data transmission, but also understanding of the features and imperfections of the commercial off-the-shelf (COTS) devices.

Fortunately, we find a perfect match between IGE and concurrent flooding, a technique with increasing popularity in the lower-power wireless networking based on concurrent transmission (CT). Riding on the tide of concurrent flooding, IGE can be easily introduced into wireless sensor networks. In concurrent flooding, senders rebroadcast received packets in strictly-synchronized time slots, with sub-microsecond synchronization accuracy. This feature of concurrent flooding is crucial for the linearity of received power. We show with controlled experiments that the linearity of power conditionally holds. After understanding the conditions for power linearity to hold, we propose a power control approach for IGE, which requires controlling the transmit powers of nodes across multiple slots to form a full-rank matrix. 
We first design controlled experiments to show that IGE is feasible via power control. After that, we integrate IGE with a recent low-power flooding protocol, BlueFlood~\cite{nahas:blueflood2021}, to demonstrate the feasibility of IGE via concurrent flooding. We believe that efficient IGE is the key enabler for the practical use of many existing network resource scheduling algorithms that assume known interference graphs.

In summary, we make the following key contributions:\\
\noindent\textbf{(1)} 
We propose to marry the challenging measurement task of interference graph estimation with the data transmission task of concurrent flooding, to conduct interference estimation simultaneously with concurrent flooding. \\
\noindent\textbf{(2)}
We conduct experiments to reveal several non-linearity issues of power to show that the linearity assumption conditionally holds under both BLE 5 PHYs and IEEE 802.15.4.\\
\noindent\textbf{(3)}
We implement IGE atop BlueFlood to demonstrate that IGE via concurrent flooding is feasible in real-world environment.

\section{Background \& Related Work}

\noindent\textbf{Interference graph estimation.} 
Traditional wisdom for network management allocates separate network resources for the measurement and data transmission tasks, resulting in the competing relation between the two types of tasks. To control the measurement overhead, cellular networks leverage their capabilities in fine-grained resource allocation and only insert the CSI-IM reference signals on a small portion of subcarriers to measure interference~\cite{cellular_reference_signal_1, cellular_reference_signal_2}. However, low-power COTS devices typically do not have such capabilities and use the whole bandwidth either for measurement, control, or data transmission~\cite{whole_bw_2, whole_bw_1}. This makes the heavy measurement tasks, e.g., IGE, not suitable for wireless sensor networks, despite that extensive works have shown the benefits of knowing the interference graph in wireless sensor networks~\cite{wsn_benefit_ig_2, wsn_benefit_ig_1}. Our work proposes to avoid the heavy measurement overhead of IGE by integrating it into a popular flooding technique in recent years.

\noindent\textbf{Concurrent flooding.}
Glossy achieves fast and efficient network flooding in a strictly-synchronized network under IEEE 802.15.4, where senders rebroadcast received packets in each slot~\cite{glossy}. A recent work, BlueFlood, extends concurrent flooding to Bluetooth 5~\cite{nahas:blueflood2019}. Due to the CFOs between concurrent senders, the receiver will see beating patterns with periodic hills (constructive interference) and valleys (destructive interference)~\cite{beating}, which are found under both Bluetooth PHYs and IEEE 802.15.4~\cite{phy_layer_replication}. Beating patterns are strong when the received signal strengths from two concurrent senders are close, causing burst errors due to deep valleys with weak signal strengths~\cite{beating_strength}. 
When the received signal strength from one sender is much larger than the rest, the beating patterns are weak and the capture effect dominates, similar to data transmission with no beating~\cite{rssispy}. Our work extends the use of concurrent flooding for IGE, which can later be used for network resource scheduling.

\noindent\textbf{Interference-aware scheduling.}
Interference graphs are extensively used for the resource management in wireless networks~\cite{interference_aware_scheduling_2, interference_aware_scheduling_1}. An efficient approach to IGE could greatly unleash the power of resource scheduling for wireless sensor networks. For example, with an interference graph, each sender can choose a proper transmit power to control its interference to others~\cite{interference_graph_power_control}, and more senders can be scheduled to transmit at the same time for better spatial reuse~\cite{interference_graph_spatial_reuse}. Moreover, CT can avoid strong destructive interference with power control given the interference graph.

\section{Interference Graph Estimation}
\subsection{Core Insight}

\begin{figure}[t!]
  \centering
  \includegraphics[scale=0.35]{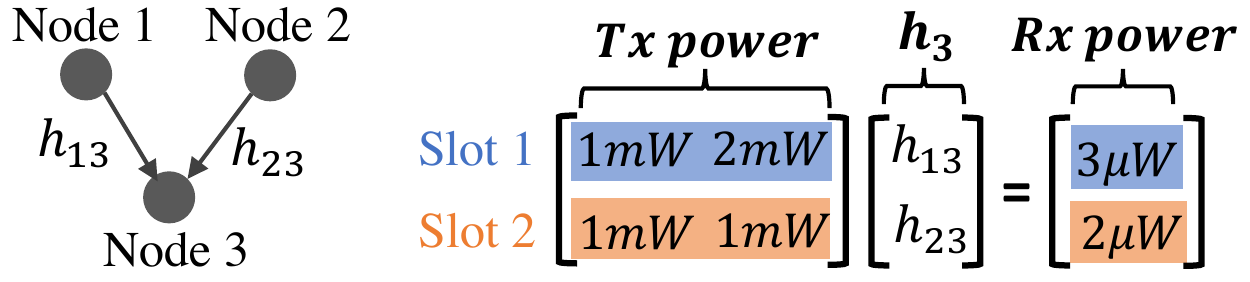}
  \caption{An example of IGE with power control}
  \label{fig:example}
\end{figure}

\noindent\textbf{An example.}
In Figure \ref{fig:example}, nodes 1 and 2 send a packet to node 3 concurrently in two time slots. In slot 1, nodes 1 and 2 send to node 3 with the transmit powers of 1mW and 2mW, respectively, and node 3 receives  at 1$\mu$W. In slot 2, nodes 1 and 2 change their transmit powers to 2mW and 1mW, respectively, and node 3 receives at 1$\mu$W. Assuming that the received power of node 3 is a linear combination of the transmit powers of nodes 1 and 2, we can have two equations for the two slots and obtain $h_{13} = h_{23} = 0.001$, where $h_{ij}$ is the channel gain from node $i$ to $j$.

\noindent\textbf{Linearity assumption of received power.} This example makes an important assumption about the linearity of received power, making it possible to estimate the channel gains by solving a system of equations. The linearity of received power is a common assumption for interference modelling in wireless networks~\cite{interference_modelling_1, interference_modelling_2}, but past measurement studies have reported different observations~\cite{additivity_hold, additivity_untrue}. We will conduct a comprehensive measurement study of the power linearity in \S\ref{sec:power_linearity}.

\noindent\textbf{Full-rank constraint.}
This example has a unique solution for the channel gains because the two equations are linearly independent. To generalize this, we want the transmit power matrix, with each row being the transmit powers of nodes in a slot, to be full-rank, such that the channel gains have a unique solution.

\subsection{Power Linearity on the COTS Devices}
\label{sec:power_linearity}

\noindent\textbf{Two properties of linearity.}
Under the linearity assumption, the received power of node $i$ from two concurrent senders can be expressed as
\begin{equation}
    p^{rx}_i = h_{1i}p^{tx}_1 + h_{2i}p^{tx}_2,
\end{equation}
where $p^{rx}_i$ and $p^{tx}_i$ are the received and transmit powers of node $i$, respectively. This assumption entails two properties of received power, both of which need to be examined on the COTS devices: 1) \textit{proportionality}, which requires the received power to be proportional to the transmit power, i.e., $p^{rx}_{i\rightarrow j} = h_{ij}p^{tx}_i$, where $p^{rx}_{i\rightarrow j}$ denotes the received power of node $j$ solely from node $i$; 2) \textit{additivity}, which requires the received powers to be additive, i.e., $p^{rx}_i = \sum_j p^{rx}_{j\rightarrow i}$.

\begin{figure}[t!]
  \centering
  \includegraphics[scale=0.45]{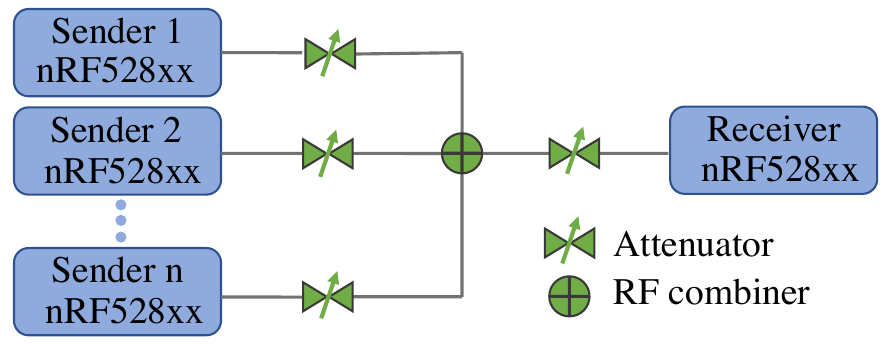}
  \caption{Experimental setup}
  \label{fig:setup}
\end{figure}

\noindent\textbf{Experimental setup.} 
We want to examine the two properties of linearity on the COTS devices. To avoid external interference, we conduct our experiments by connecting all nodes with cables. Figure \ref{fig:setup} shows the setup with multiple concurrent senders and one receiver, where the attenuators are used to control the received power from each sender at a granularity of 1 dB, and the RF combiner is to mix signals from the concurrent senders. Depending on the specific experiments, different numbers of concurrent senders may be used.
All nodes are equipped with the Nordic nRF52 series SoCs to study both Bluetooth 5 and IEEE 802.15.4. Each node sends a packet of 255 bytes to allow sufficient time for the receiver to take RSSI samples. With a sampling rate of $10^6$ samples per second, the receiver collects 1,200 RSSI samples continuously and calculates the average as the received power.

\begin{figure}[t!]
  \centering
  \includegraphics[scale=0.45]{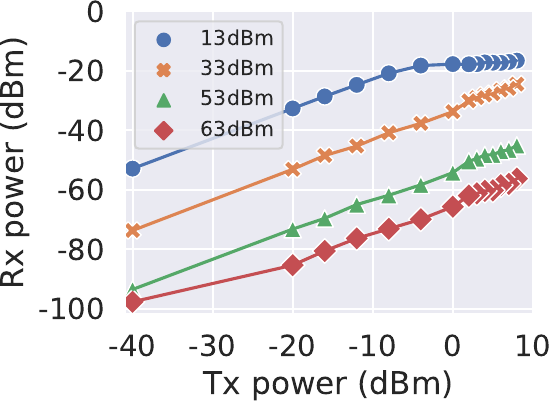}
  \caption{Proportionality between Tx and Rx powers}
  \label{fig:proportionality}
\end{figure}

\noindent\textbf{Proportionality between Tx and Rx powers.} 
We expect the received power to be proportional to the transmit power, while hardware imperfections may affect the proportionality. 
To examine this, we need one sender and one receiver for experiments. Figure \ref{fig:proportionality} shows the received powers under different transmit powers and attenuations for nRF52840 SoCs, where the cable and RF combiner together contribute about 3dB attenuation. We can see that the received power increases by almost the same amount in dB as the transmit power does in the \emph{linear region}, where the transmit power is below or equal to 0dBm and the received power is between -90dBm and -20dBm. The linear region closely matches the valid operating range (-90dBm to -20dBm) for the received power of nRF52 series, and nonlinearity starts to become significant when the received power is below $\mbox{-90}$dBm. We are surprising to find that when the transmit power is greater than 0dBm, it severely deviates from the linear relation with the corresponding received power. Since there is a sudden jump in deviation for the transmit powers greater than 0dBm, the culprit is likely to be the inaccuracy of the transmit power. 


\noindent\textbf{Additivity of received power.} 
We next want to know if the additivity assumption of received power is true on the COTS devices. We start by checking this assumption under the simple case of two concurrent senders and one receiver. 
The metric to measure power linearity, referred to as the \emph{power ratio}, is defined as the ratio of the actual received power to the sum of individually attenuated powers from the senders. A power ratio close to one indicates strong additivity, and the additivity weakens when the power ratio deviates from one. 
We conduct experiments with transmit and received powers in the linear region to exclude the disproportionality issue above. For nRF52 series SoCs, there are only 7 choices for the transmit powers below or equal to 0 dBm. We create granular power deltas between the received powers with adjustable attenuators. To simulate the case that senders in concurrent flooding forward the same packet, we simply ask the two senders to transmit all ones.


\begin{figure*}[t!]
\centering
    \subfigure[CDF]
    {\label{fig:cdf_power_ratio}
		\includegraphics[scale=0.6]{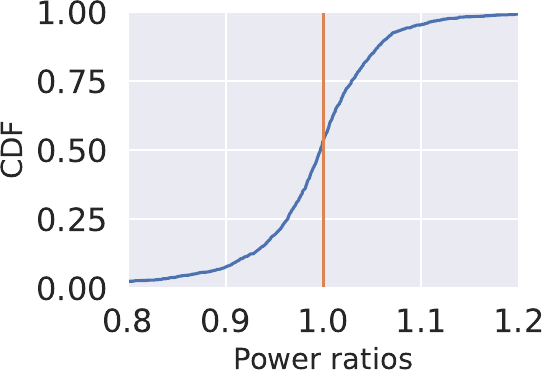}}
    \subfigure[Power ratios under strong received powers]
    {\label{fig:heat_map_ble_high_region}
		\includegraphics[scale=0.6]{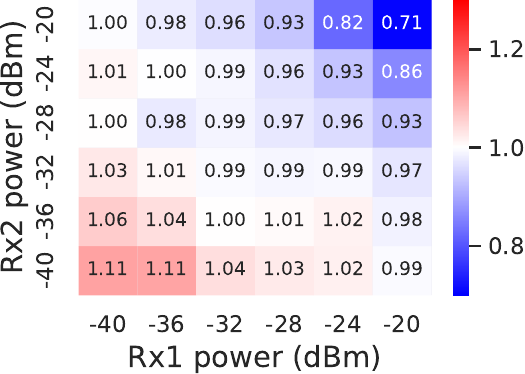}}
    \subfigure[Power ratios under weak received powers]
    {\label{fig:heat_map_ble_low_region}
		\includegraphics[scale=0.6]{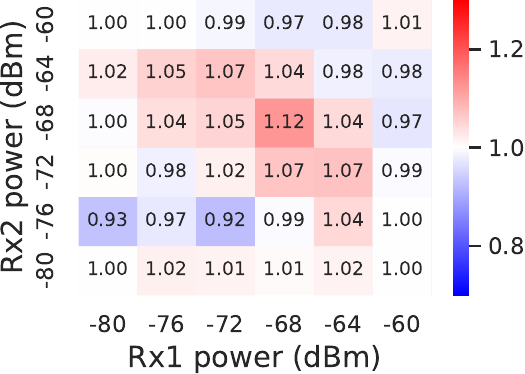}}
    \caption{Conditions for the additivity of received powers}
    \label{fig:conditions_additivity}
\end{figure*}

Figure \ref{fig:cdf_power_ratio} shows the distribution of power ratios, where about 88\% of power ratios fall within the range between 0.9 and 1.1. 
Since the operating range of the received power is critical to power linearity as discussed before, we next relate power ratios with the received power.
Figures \ref{fig:heat_map_ble_high_region} and \ref{fig:heat_map_ble_low_region} show the power ratios under the strong (-40dBm to -20dBm) and weak (-80dBm to -60dBm) received powers, respectively, where the received powers are divided into equal-length bins of 4 dB and the centers of the bins are used as labels for the $x$- and $y$-axes. We can see that the power ratio highly depends on the strength of received powers in the region of strong received powers, where the average power ratio gradually decreases from 1.11 to 0.71 when the received powers both increase from -40dBm to -20dBm along the diagonal. The additivity of received powers is the best when both the received powers are centered around $\mbox{-32}$dBm or when the received powers differ significantly from each other. The power ratio gradually approaches to 1 as the power delta between received powers increases, because the larger received power dominates. 
For weak received powers, except for the region near (-68dBm, -68dBm), all other regions have power ratios close to 1. In general, apart from the strong received powers greater than -24dBm, weak and strong received powers demonstrate similar addivitity. 

\subsection{Full-rank constraint}
Suppose that there are $n$ senders in a time-slotted multi-hop network. Each node can measure the received signal strength in a time slot it does not transmit data. Let the transmit and received powers of node $i$ at the $j$-th slot be denoted as $p_i^{tx}[j]$ and $p_i^{rx}[j]$, respectively. The channel gain from node $i$ to node $j$ is $h_{ij}$, and the channel gain vector, $\bm{h}_i = [h_{1i},\dots,h_{ni}]^T \in \mathbb{R}^{n}$, includes the channel gains from all senders to node $i$. At a time slot, if the power linearity holds, we can write the received power as a linear combination of the channel gains and the transmit powers, i.e., $p_i^{rx}[j] = \sum_{z=1}^n h_{zi}p_z^{tx}[j]$. We assume that the channel coherence time is much larger than the slot time. By varying the transmit powers of senders across time slots, we can obtain a different received power for node $i$ in each slot.
Let $\bm{p}_i^{rx} \in \mathbb{R}^{m}$ be the received powers of node $i$ in $m$ time slots, and $\mathbf{P} \in \mathbb{R}^{m\times n}$ be the corresponding transmit power matrix of the senders, where each row of the matrix dictates the transmit powers of senders in a time slot. We can have that
\begin{equation}
    \bm{p}_i^{rx} = \mathbf{P}\bm{h}_i.
\end{equation}
If the rank of the transmit power matrix, rank($\mathbf{P}$), is equal to $n$, i.e., $\mathbf{P}$ is \emph{full-rank}, we can obtain a unique solution for $\bm{h}_i$. In order for rank($\mathbf{P}$) to be greater than $n$, the number of time slots, $m$, should be greater than or equal to $n$.

\subsection{IGE: Controlled Experimental Study}
\label{sec:controlled_ige}

Before discussing how to achieve IGE via CT-based flooding, we first want to validate with experiments that IGE is feasible on the COTS devices. To this end, we use five senders and one receiver in our setup, where the senders are synchronized by a short synchronization packet from the receiver and transmit packets with the BLE 1M mode, as before. To construct a full-rank transmit power matrix, we ask each sender to pick a random transmit power in the linear region for each transmission after the time is synchronized. This process is repeated until we have sufficient measurements for each combination of  transmit powers. Let the $i$-th combination be $\bm{c}_i = [p^{tx}_1[i], \dots, p^{tx}_n[i]]^T$. We need at least $n$ combinations to construct a full-rank matrix for $n$ senders.
We calculate the \emph{channel gain error} as $|\hat{h}_{ij} - h_{ij}|$, where $\hat{h}_{ij}$ is the estimated channel gain for $h_{ij}$ in dB.

\begin{figure}[t!]
\centering
    \subfigure[Impact of condition numbers]
    {\label{fig:condition_number}
		\includegraphics[scale=0.4]{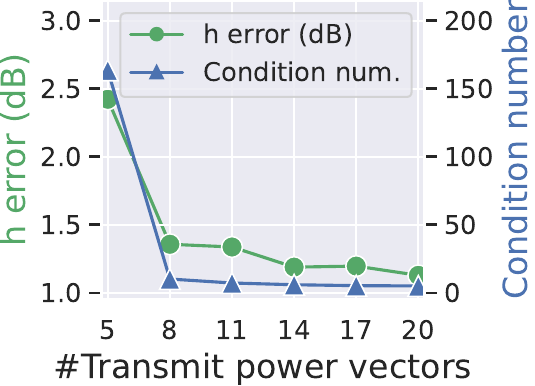}}
    \subfigure[CDF of channel gain errors]
    {\label{fig:h_cdf}
		\includegraphics[scale=0.45]{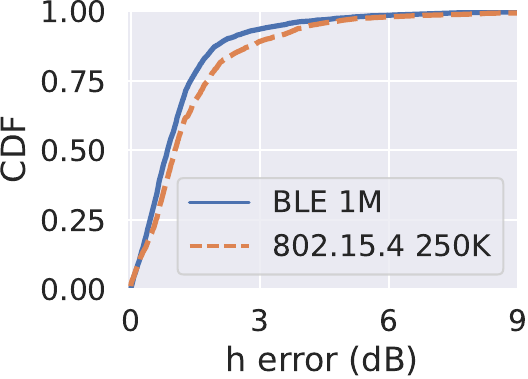}}
    \caption{Feasibility of interference graph estimation on the COTS devices}
    \label{fig:ige}
\end{figure}

With the measurement data above, we first want to understand if averaging multiple measurements for the same combination of transmit powers helps the channel gain estimation. We calculate the channel gain errors when different sample sizes are used to calculate the average transmit powers for each combination. There is no significant improvement for using a larger sample size, which implies that our measured received powers for the same node are stable. We continue this experiment using single samples of received powers to estimate channel gains. Based on the perturbation theory for linear systems, we know that the estimation error of channel gains, $\hat{h}_{ij}$, depends on the condition number of $\mathbf{P}$~\cite{perturbation}. We thus estimate the channel gains using different combinations of transmit powers (or \emph{transmit power vectors}) and obtain a relation between the estimation error and the condition number. We can see from Figure \ref{fig:condition_number} that the average channel gain error decreases together with the condition number. When the total number of transmit power vectors forming the full-rank matrix is more than 11, the condition number starts to decrease at an extremely slow rate. This agrees with the perturbation theory in that a large condition number amplifies the measurement errors and results in a large estimation error. When choosing among candidate transmit power matrices, we prefer the one with a smaller condition number.

The perturbation also suggests that it is more difficult to estimate smaller channel gains. To verify this, we use the adjustable attenuators to create different combinations of channel gains from the five senders to the receiver. In total, 30 combinations of channel gains are created and each is estimated using a full-rank matrix. Based on Figure \ref{fig:condition_number}, we use 11 transmit power vectors for the channel gain estimation. Our experimental results show that the largest channel gains experience the least estimation errors and that more than 90\% of channel gain errors are less than 3 dB for both Bluetooth and 802.15.4 (Figure \ref{fig:h_cdf}). This means that if the available transmit powers are more than 3dB apart, the estimated channel gains can then be used to choose the best transmit powers with a very high probability.

\section{Efficient IGE via CT-based Flooding}
This section presents how to integrate IGE with CT-based flooding such that interference graph can be estimated simultaneously with flooding.

\subsection{Flooding with IGE}

\begin{figure}[t!]
  \centering
  \includegraphics[scale=0.4]{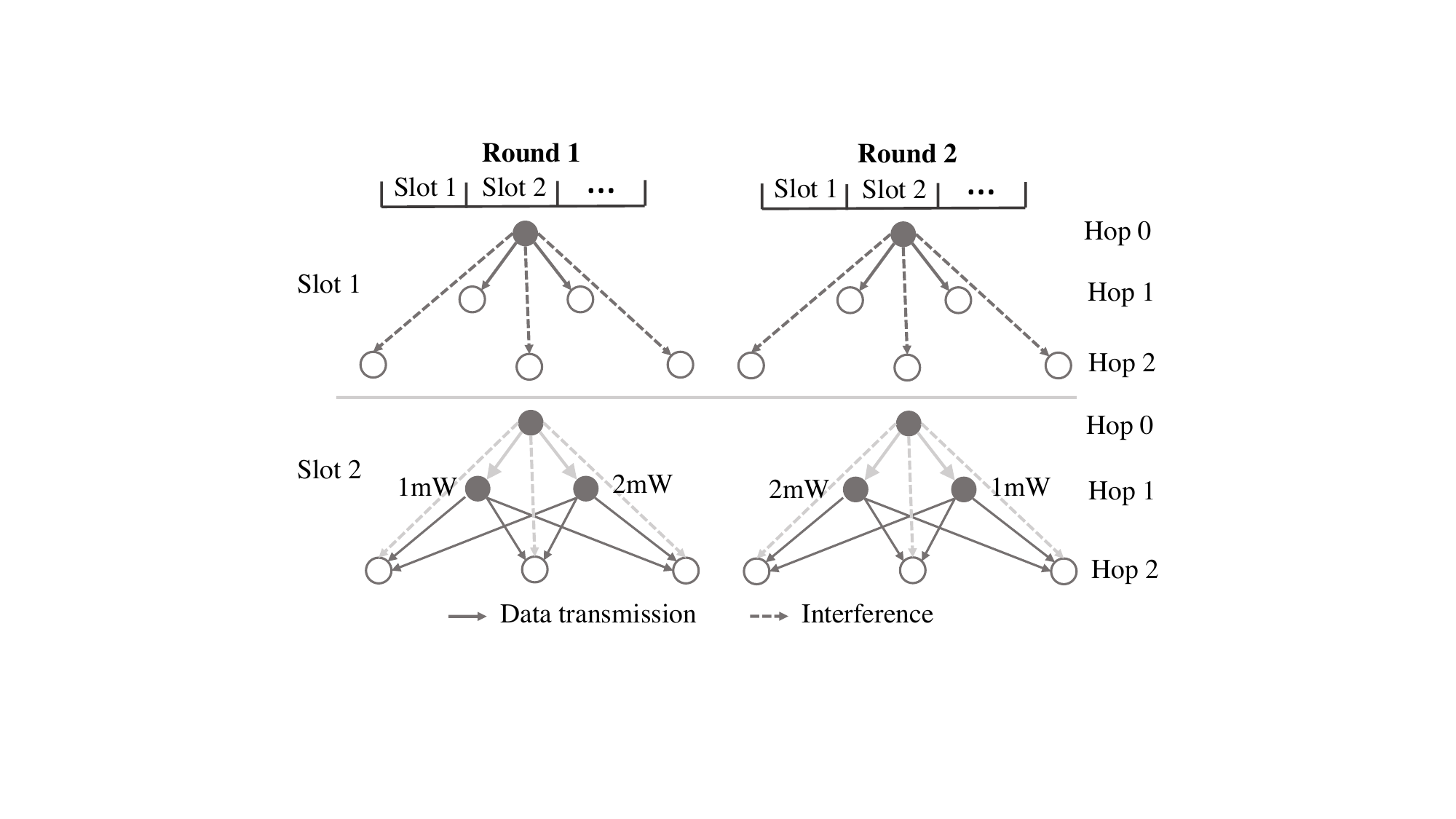}
  \caption{An example of flooding with IGE}
  \label{fig:flooding_w_ige}
\end{figure}

In a time-slotted multi-hop network, flooding requires each node to retransmit the received packet for the following $N_{tx}$ consecutive time slots, where $N_{tx}$ is the required number of retransmissions. Like Glossy, we assume that flooding is conducted round by round, each including a fixed number of slots. Figure \ref{fig:flooding_w_ige} shows an example of flooding with IGE in a two-hop wireless network, where the senders and receivers are represented by the solid and empty circles, respectively. In slot 1, the initiator broadcasts a packet to all the rest nodes. Hop-1 nodes are within the communication range of the initiator and thus can receive the packet, while hop-2 nodes can only measure the weak signal strength from the initiator for being beyond the communication range. In slot 2, hop-1 nodes have received the packet from the initiator and begin to rebroadcast it, together with the initiator. We assume that the initiator uses the same transmit power for both slots 1 and 2. Based on the additivity of received power, since hop-2 nodes have measured the interference from the initiator in slot 1, by subtracting the interference from the total received power in slot 2, they can obtain the strength of signals solely from hop-1 nodes. This implies that nodes in the same hop can vary their transmit powers across rounds to form a full-rank matrix for estimating the channel gains from themselves to their next hops. In this example, the two hop-1 nodes transmit at 1mW and 2mW in round 1, respectively, and simply switch their transmit powers in round 2. This constructs a 2$\times$2 full-rank matrix for estimating the channel gains from hop-1 nodes to hop-2 nodes.

\begin{figure}[t!]
  \centering
  \includegraphics[scale=0.45]{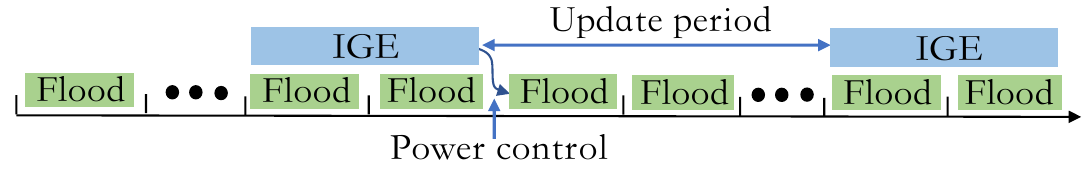}
  \caption{Interference graph update for flooding}
  \label{fig:IGE_and_flooding}
\end{figure}

From the above example, we know that by carefully controlling the transmit powers of nodes in the same hop, we can estimate the channel gains from these nodes to their next hops. As power control does not interfere the flooding process, we can achieve IGE simultaneously with flooding. As shown in Figure \ref{fig:IGE_and_flooding}, interference graph is updated periodically to adapt to the changing network conditions, where the \emph{update period} determines how often the interference graph should be updated. It is noticeable that calculating channel gains requires knowing the received powers of nodes that are locally measured. We can simply assume that the calculation of channel gains is conducted in a centralized way at the initiator and that each node needs to report their local measurements back to the initiator during the IGE process. After a plan for power control is created, the initiator disseminates it to all the nodes and each node will adjust its transmit power accordingly. To reduce communication overhead, the plan for each node can be simply represented by a number, i.e., the index of a set of pre-defined options, and the data collection and dissemination for IGE can be piggybacked onto the normal network traffic, e.g., flooding and converagecast. 

\section{Flooding over Bluetooth}
\label{sec:impl}

This section presents and evaluates the implementation of our approach based on BlueFlood \cite{nahas:blueflood2021}, a recent lower-power flooding protocol implemented on the Nordic nRF SoCs. We use Nordic nRF52840 nodes for system implementation and experiments. 

\subsection{Experimental Setup}
\noindent\textbf{Implementation.} In our implementation, we modify Blueflood to make each node capable of adjusting its configurations according to a scheduling plan generated by a central node, called the \emph{initiator}. This scheduling plan specifies how each node should adjust its transmit power in each time slot. The transmit powers of all nodes in the same time slot are represented as a transmit power vector, where the same transmit power vector is used for the entire round. Suppose that $m$ transmit power vectors are used for IGE. This requires each node to follow a sequence of $m$ transmit powers spanning $m$ rounds for data transmission. We require the duration of IGE to be multiples of $m$ rounds such that the received powers of nodes under the same transmit power can be measured multiple times, to mitigate the randomness in single measurements. During IGE, each node measures the average received power in each round when it is in listen mode and sends back the estimates to the initiator. The initiator will then obtain the received powers of nodes. Considering the limited computing resources of the initiator, we currently connect the initiator to a desktop, which estimates the channel gains and determines the transmit powers of nodes. The received power is estimated as the average of received RSSI samples.

\begin{figure}[t!]
  \centering
  \includegraphics[scale=0.3]{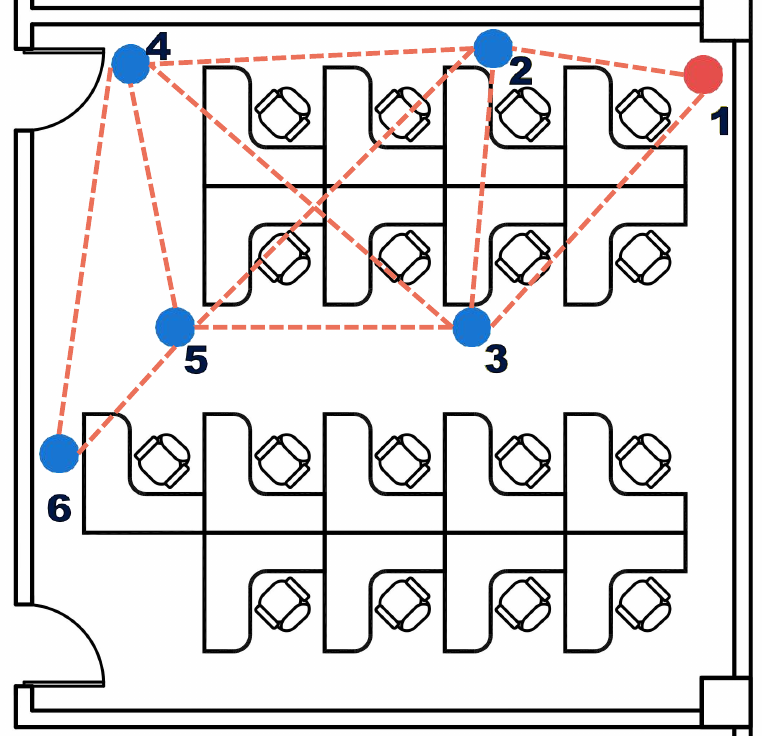}
  \caption{Our testbed}
  \label{fig:topology}
  \vspace{-1em}
\end{figure}

\noindent\textbf{Testbed and configuration.} We run experiments in a testbed deployed in our office, with an area of 10m$\times$10m. The office layout and the locations of nodes are shown in Figure~\ref{fig:topology}. The deployed network is of three hops with node $1$ being the initiator.
To demonstrate how our approach can effectively mitigate the destructive interference due to CTs, we deliberately choose the channel gains from nodes 4 and 5 to node 6 to be close and have a difference within 1dB. To mitigate the impact of external interference, we enforce flooding to use only a single channel, specifically, the Bluetooth advertising channel 37. Other parameter settings are the same as what BlueFlood used in~\cite{nahas:blueflood2021}, where a transmission-reception ratio of $N_{Tx} = 3$ and a packet size of 38 bytes are used. 

\subsection{Performance Evaluation}

We want to evaluate the accuracy of IGE in real-world settings, to show that IGE with power control is feasible.

\noindent\textbf{Accuracy of IGE.}
As flooding is from the initiator to downstream nodes, we only consider the downstream links from each node to its next hops. To conduct IGE together with flooding, we simply construct full-rank transmit power matrices by shuffling available transmit powers in the linear region, such that transmit power vectors across rounds are different. In our network, the maximum of nodes in a hop is 2, indicating that we need at least two transmit power vectors. To understand the accuracy of IGE, we want to compare the estimated channel gains from the flooding-based IGE against those from the point-to-point channel estimation, which is used as the ground-truth. The impact of channel variations is mitigated by conducting the flooding-based and point-to-point IGEs back to back. We repeat this back-to-back experiments until we have 1500 rounds of measurement data, where each IGE lasts for 30 rounds. 

\begin{figure}[t!]
\centering
    \subfigure[Distribution of estimation errors]
    {\label{fig:h_error_cdf_testbed}
		\includegraphics[scale=0.43]{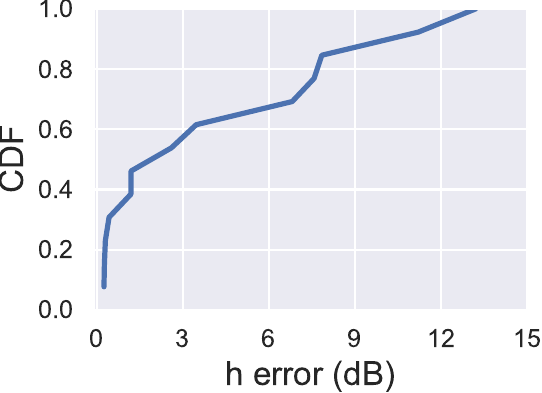}}
    \subfigure[Distribution of small $h$'s]
    {\label{fig:h_max_ratio_cdf}
		\includegraphics[scale=0.43]{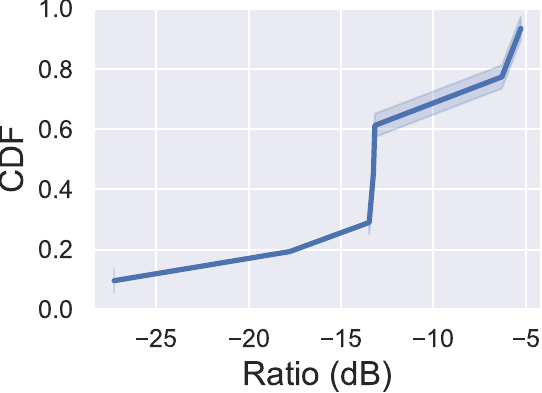}}
    \subfigure[Impact of condition number]
    {\label{fig:impact_cond_number_testbed}
		\includegraphics[scale=0.40]{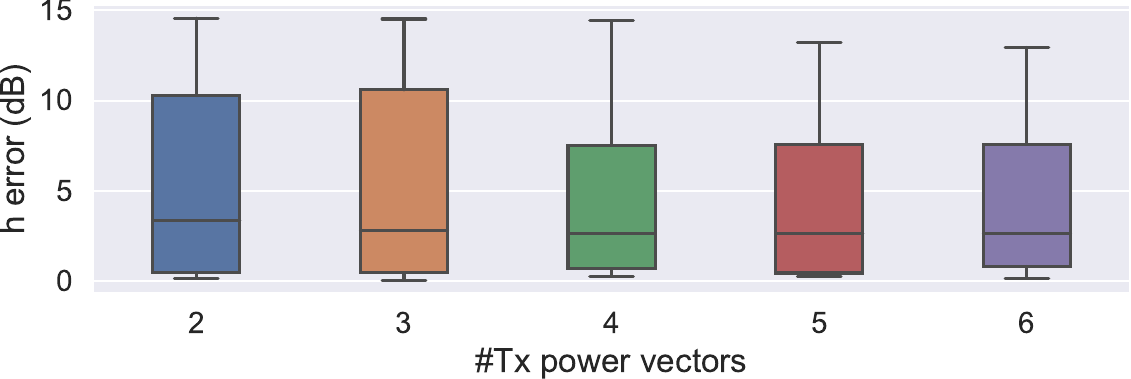}}
    \caption{Accuracy of IGE on our testbed}
\end{figure}


Figure \ref{fig:h_error_cdf_testbed} shows the distribution of channel gain errors for flooding-based IGE when four transmit power vectors are used. We can see that 60\% of the estimated channel gains have an error less than 3dB. For channel gains with an error greater than 3dB, we compute its relative magnitude with the maximum channel gain in the network and find that 70\% of the small channel gains are less than the maximum one for more than 10dB, which significantly mitigates the impact of large estimation errors. Compared with the controlled experiments in \S\ref{sec:controlled_ige}, we find the channel gain errors are larger in real-world environment due to varying network conditions. To understand if using more transmit power vectors helps increase estimation accuracy, we re-do the experiments above using different numbers of transmit powers. Figure \ref{fig:impact_cond_number_testbed} shows the boxplot of channel channel gain errors when two to six transmit power vectors are used. It can be seen that the range of estimation errors decreases sharply when the number of transmit power vectors increases from three to four. By examining the condition numbers of the $4\times 2$ transmit power matrices used in experiments, we find that the average condition number is as small as 1.48, very close 1. Based on the perturbation theory for linear systems, we know that the choice of transmit power matrices in our experiments does not significantly amplify the measurement errors and is not a major source of error. Our experimental results show that IGE via concurrent flooding in real-world environment experiences larger estimation errors than in controlled experiments, but is still feasible, where the channel gains with large estimation errors are much smaller than those with small estimation errors.

\section{Discussions \& Limitations}
\nonumber\textbf{Nonlinearity issues.}
Even though nearly 90\% of the power ratios in our experiments fall between 0.9 and 1.1, but there are still some outliers beyond this range. We have spotted several nonlinearity issues of power in the COTS devices in \S\ref{sec:power_linearity}, but there may exist more issues to discover. However, the limited visibility into the hardware of nRF52 series SoCs may prevent us from diving deeper into the nonlinearity issues. The nonlinearity issues weaken the linear relation between transmit and received powers and are thus major sources of errors for the accuracy of IGE.

\nonumber\textbf{Overhead of IGE and optimization.} Although our approach achieves IGE and flooding using the same frequency-time resources, both the data collection and dissemination for IGE incur communication overhead. We believe that the communication overhead can be efficiently controlled with several approaches: 1) reducing the frequency of IGE or only conducting IGE when needed, 2) designing an efficient representation for control and measurement data, and 3) piggybacking these data onto normal traffic. Besides, estimating interference graphs also consumes computing resources. We have done this with a destkop connected to the initiator, which may not be necessary if the estimation can be conducted efficiently within the low-power devices.

\nonumber\textbf{Broader impact of the interference graph.}
We have demonstrated how to achieve IGE together with CT-based flooding. In fact, as interference graphs are widely used in wireless network resource management, the interference graph estimated from concurrent flooding can also be used to benefit many other network activities. Considering the multi-faceted contributions of IGE to the network, the related overhead for measurement is worthwhile.

\section{Conclusions}
In this paper, we proposed to integrate interference graph estimation into data transmission tasks such that it can be done simultaneously with the data transmission tasks using the same frequency-time resources. We showed that interference graph estimation on the Nordic nRF52 series SoCs was feasible in both controlled and real-world experiments, where the real-world experiments were done by integrating interference graph estimation atop a recent low-power flooding protocol. Our experimental results showed that concurrent flooding is a perfect carrier for this integration in wireless sensor networks. As for future work, we want to design network protocols for collecting the RSSI measurements and disseminating the plans for power control.

%
%
\section{Acknowledgments}
We appreciate the constructive feedback from the anonymous reviewers. We thank our colleagues, Aimin Tang and Xuyang Lu, for their valuable discussion and feedback on this paper. This work was supported by the National Natural
Science Foundation of China under Grant 62201346.


%
%
\balance
\bibliographystyle{abbrv}
\bibliography{sigproc}  

@INPROCEEDINGS{ige_traditional_wisdom,
	AUTHOR = "Zhou, G. and He, T. and Stankovic, J.A. and Abdelzaher, T.",
	TITLE = "{RID}: {R}adio interference detection in wireless sensor networks.",
	BOOKTITLE = "Proceedings IEEE 24th Annual Joint Conference of the IEEE Computer and Communications Societies",
	PAGES = "891-901",
	MONTH = "March", 
	YEAR = {2005}	}

@INPROCEEDINGS{additivity_hold,
	AUTHOR = "Maheshwari, R. and Jain, S. and Das, S. R.",
	TITLE = "A measurement study of interference modeling and scheduling in low-power wireless networks",
	BOOKTITLE = "Sensys",
	PAGES = "141-154",
	YEAR = {2009}}

@INPROCEEDINGS{glossy,
	AUTHOR = "F. Ferrari and M. Zimmerling and L. Thiele and O. Saukh",
	TITLE = "Efficient network flooding and time synchronization with {G}lossy",
	BOOKTITLE = "IPSN",
	PAGES = "73-84",
	YEAR = {2011}}

@INPROCEEDINGS{beating,
	AUTHOR = "Liao, C.H. and Katsumata, Y. and Suzuki, M. and Morikawa, H.",
	TITLE = "Revisiting the so-called constructive interference in concurrent transmission",
	BOOKTITLE = "LCN",
	PAGES = "280-288",
	YEAR = {2016}}

@INPROCEEDINGS{beating_strength,
	AUTHOR = "Baddeley, M. and Boano, C.A. and Escobar-Molero, A. and Liu, Y. and Ma, X. and Raza, U. and Römer, K. and Schuß, M. and Stanoev, A.",
	TITLE = "The impact of the physical layer on the performance of concurrent transmissions",
	BOOKTITLE = "ICNP",
	PAGES = "1-12",
	YEAR = {2020}}

@INPROCEEDINGS{phy_layer_replication,
	AUTHOR = "Jacob, R. and Schaper, A.B. and Biri, A. and Da Forno, R. and Thiele, L.",
	TITLE = "Synchronous transmissions on {B}luetooth 5 and {IEEE} 802.15.4–{A} replication study",
	BOOKTITLE = "3rd Workshop on Benchmarking Cyber-Physical Systems and Internet of Things (CPS-IoTBench 2020)",
	YEAR = {2020}}

@INPROCEEDINGS{additivity_untrue,
	AUTHOR = "Son, D. and Krishnamachari, B. and Heidemann, J.",
	TITLE = "Experimental study of concurrent transmission in wireless sensor networks",
	BOOKTITLE = "Sensys",
	PAGES = "237-250",
	YEAR = {2006}}

@INPROCEEDINGS{wsn_interference_2,
	AUTHOR = "Li, X.-Y. and Moaveni-Nejad, K. and Song, W.-Z. and Wang, W.-Z.",
	TITLE = "Interference-aware topology control for wireless sensor networks",
	BOOKTITLE = "SECON",
	PAGES = "263-274",
	YEAR = {2005}}

@INPROCEEDINGS{cellular_reference_signal_1,
	AUTHOR = "Elgendi, H. and Mäenpää, M. and Levanen, T. and Ihalainen, T. and Nielsen, S. and Valkama, M.",
	TITLE = "Interference measurement methods in {5G NR}: {P}rinciples and performance",
	BOOKTITLE = "2019 16th International Symposium on Wireless Communication Systems (ISWCS)",
	PAGES = "233-238",
	YEAR = {2019}}

@INPROCEEDINGS{wsn_benefit_ig_2,
	AUTHOR = "Chen, S. and Coolbeth, M. and Dinh, H. and Kim, Y.-A. and Wang, B.",
	TITLE = "Data Collection with Multiple Sinks in Wireless Sensor Networks",
	BOOKTITLE = "WASA",
	PAGES = "284-294",
	YEAR = {2009}}

@INPROCEEDINGS{rssispy,
	AUTHOR = "C. Herrmann and M. Zimmerling",
	TITLE = "{RSSISpy}: {I}nspecting Concurrent Transmissions in the Wild",
	BOOKTITLE = "EWSN",
	YEAR = {2022}}

@INPROCEEDINGS{cellular_reference_signal_2,
	AUTHOR = "Peralta, E. and Pocovi, G. and Kuru, L. and Jayasinghe, K. and Valkama, M.",
	TITLE = "Outer Loop Link Adaptation Enhancements for Ultra Reliable Low Latency Communications in {5G}",
	BOOKTITLE = "2022 IEEE 95th Vehicular Technology Conference (VTC2022-Spring)",
	PAGES = "1-7",
	YEAR = {2022}}

@ARTICLE{interference_aware_scheduling_1,
	AUTHOR = "Yu, Y. and Dutkiewicz, E. and Huang, X. and Mueck, M.",
	TITLE = "Downlink resource allocation for next generation wireless networks with inter-cell interference",
	JOURNAL = {IEEE Transactions on Wireless Communications},
	VOLUME = {12},
	NUMBER = {4},
	PAGES = {1783-1793},
	YEAR = {2013}	}

@ARTICLE{interference_aware_scheduling_2,
	AUTHOR = "Cao, J. and Peng, T. and Liu, X. and Dong, W. and Duan, R. and Yuan, Y. and Wang, W. and Cui, S.",
	TITLE = "Resource allocation for ultradense networks with machine-learning-based interference graph construction",
	JOURNAL = {IEEE Internet of Things Journal},
	VOLUME = {7},
	NUMBER = {3},
	PAGES = {2137-2151},
	YEAR = {2019}	}

@ARTICLE{wsn_interference_3,
	AUTHOR = "Chen, S. and Huang, M. and Tang, S. and Wang, Y.",
	TITLE = "Capacity of data collection in arbitrary wireless sensor networks",
	JOURNAL = {IEEE Transactions on Parallel and Distributed Systems},
	VOLUME = {23},
	NUMBER = {1},
	PAGES = {52-60},
	YEAR = {2011}	}

@ARTICLE{interference_graph_power_control,
	AUTHOR = "M. M. Halldórsson",
	TITLE = "Wireless scheduling with power control",
	JOURNAL = {ACM Transactions on Algorithms},
	VOLUME = {9},
	NUMBER = {1},
	PAGES = {1-20},
	YEAR = {2012}	}

@ARTICLE{interference_graph_spatial_reuse,
	AUTHOR = "Chen, X. and Huang, J.",
	TITLE = "Distributed spectrum access with spatial reuse",
	JOURNAL = {IEEE Journal on Selected Areas in Communications},
	VOLUME = {31},
	NUMBER = {1},
	PAGES = {593-603},
	YEAR = {2013}	}

@ARTICLE{ige_spatial_dl,
	AUTHOR = "Cui, W. and Shen, K. and Yu, W.",
	TITLE = "Spatial deep learning for wireless scheduling",
	JOURNAL = {IEEE Journal on Selected Areas in Communications},
	VOLUME = {37},
	NUMBER = {6},
	PAGES = {1248-1261},
	YEAR = {2019}	}

@ARTICLE{ige_embedding,
	AUTHOR = "Lee, M. and Yu, G. and Li, G.Y.",
	TITLE = "Graph embedding-based wireless link scheduling with few training samples.",
	JOURNAL = {IEEE Transactions on Wireless Communications},
	VOLUME = {20},
	NUMBER = {4},
	PAGES = {2282-2294},
	YEAR = {2020}	}

@ARTICLE{ige_traditional_survey,
	AUTHOR = "Baccour, N. and Koubâa, A. and Mottola, L. and Zúñiga, M. A. and Youssef, H. and Boano, C. A. and Alves, M.",
	TITLE = "Radio link quality estimation in wireless sensor networks: A survey",
	JOURNAL = {ACM Transactions on Sensor Networks},
	VOLUME = {8},
	NUMBER = {4},
	PAGES = {1-33},
	YEAR = {2012}	}

@ARTICLE{interference_modelling_1,
	AUTHOR = "F. Fernandes and A. Ashikhmin and T. L. Marzetta",
	TITLE = "Inter-cell interference in noncooperative {TDD} large scale antenna systems",
	JOURNAL = {IEEE Journal on Selected Areas in Communications},
	VOLUME = {31},
	NUMBER = {2},
	PAGES = {192-201},
	YEAR = {2013}	}

@ARTICLE{whole_bw_1,
	AUTHOR = "Tang, Q. and Yang, L. and Giannakis, G. B. and Qin, T.",
	TITLE = "Battery power efficiency of PPM and FSK in wireless sensor networks",
	JOURNAL = {IEEE Transactions on wireless Communications},
	VOLUME = {6},
	NUMBER = {4},
	PAGES = {1308-1319},
	YEAR = {2007}	}

@ARTICLE{whole_bw_2,
	AUTHOR = "Main, E. and Coffing, D.",
	TITLE = "An FSK demodulator for Bluetooth applications having no external components",
	JOURNAL = {IEEE Transactions on Circuits and Systems II: Analog and Digital Signal Processing},
	VOLUME = {49},
	NUMBER = {6},
	PAGES = {373-378},
	YEAR = {2022}	}

@ARTICLE{wsn_benefit_ig_1,
	AUTHOR = "Ergen, S.C. and Varaiya, P.",
	TITLE = "{TDMA} scheduling algorithms for wireless sensor networks",
	JOURNAL = {Wireless networks},
	PAGES = {985-997},
	YEAR = {2010}	}

@ARTICLE{interference_modelling_2,
	AUTHOR = "Makarfi, A. U. and Rabie, K. M. and Kaiwartya, O. and Adhikari, K. and Nauryzbayev, G. and Li, X. and Kharel, R.",
	TITLE = "Toward physical-layer security for {I}nternet of vehicles: {I}nterference-aware modeling",
	JOURNAL = {IEEE Internet of Things Journal},
	VOLUME = {8},
	NUMBER = {1},
	PAGES = {443-457},
	YEAR = {2020}	}

@ARTICLE{wsn_interference_1,
	AUTHOR = "Harbin, J. and Burns, A. and Davis, R. I. and Indrusiak, L. S. and Bate, I. and Griffin, D.",
	TITLE = "The airtight protocol for mixed criticality wireless {CPS}",
	JOURNAL = {ACM Transactions on Cyber-Physical Systems},
	VOLUME = {4},
	NUMBER = {2},
	PAGES = {1-28},
	YEAR = {2019}	}

@ARTICLE{perturbation,
	AUTHOR = "Luo, Z.Q. and Tseng, P.",
	TITLE = "Perturbation analysis of a condition number for linear systems",
	JOURNAL = {SIAM Journal on Matrix Analysis and Applications},
	VOLUME = {15},
	NUMBER = {2},
	PAGES = {636-660},
	YEAR = {1994}	}

@inproceedings{nahas:blueflood2019,
	title = {Concurrent Transmissions for Multi-Hop Bluetooth 5},
	pages = {130--141},
	booktitle = {EWSN'19},
	author = {Al Nahas, Beshr and Duquennoy, Simon and Landsiedel, Olaf},
	urldate = {2023-01-13},
	date = {2019-03-15},
        year = 2019
}

@article{nahas:blueflood2021,
	title = {{BlueFlood}: Concurrent Transmissions for Multi-hop Bluetooth 5 — Modeling and Evaluation},
	volume = {2},
	issn = {2691-1914},
	url = {https://doi.org/10.1145/3462755},
	doi = {10.1145/3462755},
	shorttitle = {{BlueFlood}},
	pages = {22:1--22:30},
	number = {4},
	journal = {{ACM} Transactions on Internet of Things},
	shortjournal = {{ACM} Trans. Internet Things},
	author = {Nahas, Beshr Al and Escobar-Molero, Antonio and Klaue, Jirka and Duquennoy, Simon and Landsiedel, Olaf},
	urldate = {2023-02-26},
	date = {2021-07-15},
        year = 2021
}
\end{document}